# VIRUS-MNIST: A BENCHMARK MALWARE DATASET

David A. Noever and Samantha E. Miller Noever

PeopleTec, Inc., Huntsville, Alabama, USA
david.noever@peopletec.com

## ABSTRACT

The short note presents an image classification dataset consisting of 10 executable code varieties and approximately 50,000 virus examples. The malicious classes include 9 families of computer viruses and one benign set. The image formatting for the first 1024 bytes of the Portable Executable (PE) mirrors the familiar MNIST handwriting dataset, such that most of the previously explored algorithmic methods can transfer with minor modifications. The designation of 9 virus families for malware derives from unsupervised learning of class labels; we discover the families with KMeans clustering that excludes the non-malicious examples. As a benchmark using deep learning methods (MobileNetV2), we find an overall 80% accuracy for virus identification by families when beneware is included. We also find that once a positive malware detection occurs (by signature or heuristics), the projection of the first 1024 bytes into a thumbnail image can classify with 87% accuracy the type of virus. The work generalizes what other malware investigators have demonstrated as promising convolutional neural networks originally developed to solve image problems but applied to a new abstract domain in pixel bytes from executable files. The dataset is available on Kaggle and Github.

## 1. INTRODUCTION

For classifying handwriting, the popularity of the Modified National Institute of Standards and Technology dataset (MNIST) continues to dominate the early exploration of new algorithms [1-3]. Its extensions to other domains have included foreign languages [4-8], medical diagnoses [9], overhead imagery [10], and retail objects [11]. With modern deep learning and convolutional neural networks, the accuracy for multiple classification challenges typically exceed 90% across all classes [12]. Recent interest in applying the same techniques to anti-virus and malware detectors [13-19] motivates the present work to score a similar formatted problem and compare the algorithmic performance with existing methods. Intel Labs and Microsoft Threat Protection Intelligence Team recently launched their static malware collaboration called STAMINA: Scalable Deep Learning Approach for Malware Classification [20]. The contribution of this short note is to reformulate the malware-image problem as a familiar MNIST variant, to generate the 9-virus clusters based on byte-similarities, and then to identify the virus family based on a grey-scale thumbnail image (32 x 32). Figure 1 shows the abstract images derived for each of the 10 classes, with "0" as the only one that is non-malicious.

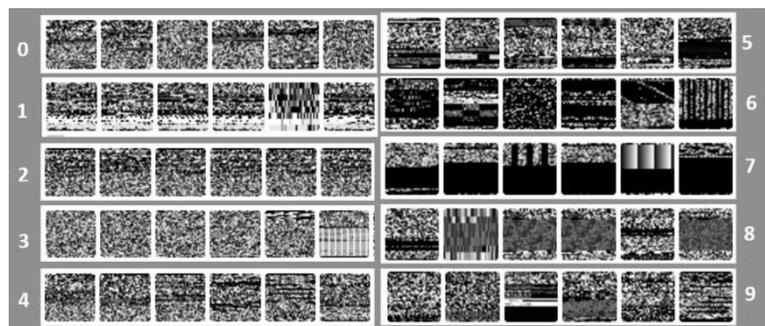

Figure 1 Virus-MNIST showing 10 classes. The "0" class represents non-malicious examples. The other 9 virus families were clustered using a K-means method to match with the standard MNIST format and multi-class

## 2. METHODS

We explore the malware dataset first combined [16] as bulk virus downloads (from virusshare.com) and non-malicious examples (from portableapps.com). Microsoft documents the format header for Portable Executable (PE) files [21-22] and importable python libraries ("pefile", [23]) exist as convenient extraction tools from compiled .exe code. The Portable Executable dataset [16] contains 51,880 examples of the first 1024 bytes (32x32 pixel values) of the header. As comma-separated-values, the entire file's MD5 hash augments the information available for each example and provides enough identifying information to trace its operating features using community-supported repositories like VirusTotal.com. Previous solutions have designated a binary class label as either malware or not (e.g. beneware) [16]. If formulated as a two-class problem, the imbalanced ratio of malware to beneware equals approximately 20:1 (49,364:2516). Without rebalancing this ratio, a 95% correct solution would simply declare all cases as malware. To recast the image dataset as an MNIST-formatted alternative, we performed a standard cluster analysis using the KMeans algorithm [24]; we assigned cluster numbers equal to 9 based on the 1024 column byte vector. We exclude the identifying file hash. We attempted to assign dominant families to each derived cluster based on the known MD5 hashes but found multiple names and sample diversity when querying VirusTotal. To reduce the number of code changes comparing a larger or smaller multi-class problem with existing MNIST infrastructure, we opted to use the assigned 9-cluster result as a fully unsupervised example. Figure 2 summarizes the class distribution following KMeans clustering for the malware class only and the 9 resulting virus families. By testing 10 or more MD5 hash values against the VirusTotal.com database, we assigned the broad types and example executable names. These choices showed sufficient diversity and overlap that the designations provide only representative choices. The outcome proves more statistical and less operational for malware behavior.

| Class | Count | Group | Type | Example |
|---|---|---|---|---|
| 0 | 2516 | Beneware | Good | putty.exe |
| 1 | 7684 | Malware | Adware | IESettings |
| 2 | 3037 | Malware | Trojan | Supreme.exe |
| 3 | 2404 | Malware | Trojan | myfile.exe |
| 4 | 796 | Malware | Installer | myfile.exe |
| 5 | 6662 | Malware | Backdoor | myfile.exe |
| 6 | 15377 | Malware | Crypto | Powershell |
| 7 | 7494 | Malware | Backdoor | BitTorrent.exe |
| 8 | 2571 | Malware | Downloader | myfile.exe |
| 9 | 3339 | Malware | Heuristic | myfile.exe |

Figure 2. Class distributions and example types for malware and beneware PE File headers.

We converted the CSV format [16] to greyscale images using the intermediate NetPBM text format (PGM) to create ASCII-raw images, then the ImageMagick [25] command-line tools for compressing the image to viewable JPEG files. We split the resulting 51,880 thumbnails into the same 85:15 ratio used by MNIST training and testing bins, such that the unseen test images help validate any algorithm's ability to generalize from the training images. The choice of 32x32 pixels to represent the PE header proves useful for later deep learning algorithms that depend on powers of 2 (in stride length) to form their convolutional layers. This differs slightly from the traditional (arbitrary) 28x28 pixel thumbnails in most MNIST variants [1-11]; the differences are cosmetic only for most algorithms other than deep learning ones. Three different formats are provided for download, including train and test sets as comma-separated values files, JPEG images sorted by class (10 total), and the original MNIST binary format (idx-ubyte) [1, 26]. These three formatting options should cover most all published MNIST solutions with only minor modifications.

For benchmarking, we trained a convolutional neural network (CNN) using transfer learning (MobileNetV2) in 50 epochs with a learning rate of 0.001. To measure the effects of malware counts in the dataset, we trained smaller (10,000) and larger (50,000) examples. MobileNetV2 [27] offers a fast learning architecture optimized for speed, accuracy, and size on mobile devices. The algorithm demonstrates an inverted residual structure with the initial fully convolution layer (32 filters), followed by 19 residual bottleneck layers. The use of residual networks provides clean, modular architectures with fewer parameters (weights) but more layers in vision applications. Transfer learning further improves recognition speeds, particularly where the weights are derived from larger, more diverse datasets like COCO and ImageNet, but only selected layers need retraining for a domain-specific dataset like malware images. In principle, the success of transfer learning captures the large visual features (like edges or blocks), then finetunes the class assignments on objects never presented in the original training cycles (like malware detection from a classifier built to recognize common household items).

## 3. RESULTS

Figure 3 shows the accuracy per class in three different CNN models. The highest accuracy across all classes averages 80%.

| Class | Count | Group | Type | Acc-1 | Acc-2 | Acc-3 |
|---|---|---|---|---|---|---|
| 0 | 2516 | Beneware | Good | 58.0% | 39.0% | 22.0% |
| 1 | 7684 | Malware | Adware | 100.0% | 99.0% | 99.0% |
| 2 | 3037 | Malware | Trojan | 81.0% | 84.0% | 91.0% |
| 3 | 2404 | Malware | Trojan | 88.0% | 90.0% | 94.0% |
| 4 | 796 | Malware | Installer | 100.0% | 100.0% | 100.0% |
| 5 | 6662 | Malware | Backdoor | 75.0% | 84.0% | 83.0% |
| 6 | 15377 | Malware | Crypto | 75.0% | 64.0% | 89.0% |
| 7 | 7494 | Malware | Backdoor | 51.0% | 72.0% | 72.0% |
| 8 | 2571 | Malware | Downloader | 68.0% | 66.0% | 76.0% |
| 9 | 3339 | Malware | Heuristic | 70.0% | 89.0% | 74.0% |

Figure 3. Accuracy per class. Acc-1 is 20% sampling, Acc-3 is 100% sampling and Acc-2 is slow learning rates.

To explore the effects of dataset size, we modeled a faster 20% sampling (Acc-1) with a learning rate of 0.001, batch size=16, and 50 epochs. A second model (Acc-2) featured a slower learning rate (0.0005) over longer training times (100 epochs). A slower learning rate can avoid disruptive steps when transfer learning from a pre-trained network like MobileNetV2. A final model (Acc-3) included 100% samples (the full 51,880 train and test sets) and a faster learning rate (0.001) and time (50 epochs). Across all 10 classes, the average accuracy varied less than 2% for the three cases but did peak at 80% for the larger dataset (Acc-3). The three models together show the highest false-negative rate for malware is class "0" or beneware that gets flagged as potentially malicious. This behavior may reflect the inexactness of the PE header as an indicator of malware, or the hijacking of benign header characteristics to disguise malware in the first 1024 bytes.

Figure 4 shows the error matrix for the first model (Acc-1), which highlights that beneware is most often mistake for backdoors (7), heuristic (9), and downloader (8). It's worth noting that particularly for a downloader that reaches out to a command and control malware site, this byte-code signature may prove malicious or benign since it depends on the website itself; a benign program may reach out to download an update from Microsoft on launch. The higher false-negative rates may also originate in the KMeans approach, given the clustering excluded the non-malicious cases to simplify the generation of 9 malware families as independent groups from beneware. Further investigation may explore whether other clustering approaches benefit the class distinctions. For instance, density-based clustering (DBSCAN) optimizes the tight groups with minimal overlap. A further enhancement would use the MD5 hash to determine virus family and avoid clustering altogether as an alternative which also prevents the CNN from simply modeling the unsupervised (KMeans) algorithm itself rather than the natural distribution of malware images.

| Actual Class \ Prediction | 0 | 1 | 2 | 3 | 4 | 5 | 6 | 7 | 8 | 9 |
|---|---|---|---|---|---|---|---|---|---|---|
| 0 | 0.59 | 0.02 | 0.03 | 0.01 | 0.00 | 0.06 | 0.10 | 0.11 | 0.04 | 0.05 |
| 1 | 0.00 | 1.00 | 0.00 | 0.00 | 0.00 | 0.00 | 0.00 | 0.00 | 0.00 | 0.00 |
| 2 | 0.11 | 0.00 | 0.81 | 0.02 | 0.00 | 0.04 | 0.00 | 0.02 | 0.01 | 0.00 |
| 3 | 0.03 | 0.00 | 0.02 | 0.88 | 0.00 | 0.00 | 0.01 | 0.05 | 0.00 | 0.00 |
| 4 | 0.00 | 0.00 | 0.00 | 0.00 | 1.00 | 0.00 | 0.00 | 0.00 | 0.00 | 0.00 |
| 5 | 0.14 | 0.00 | 0.01 | 0.00 | 0.00 | 0.75 | 0.02 | 0.02 | 0.03 | 0.02 |
| 6 | 0.16 | 0.00 | 0.00 | 0.00 | 0.00 | 0.02 | 0.75 | 0.01 | 0.02 | 0.03 |
| 7 | 0.38 | 0.00 | 0.01 | 0.01 | 0.00 | 0.02 | 0.05 | 0.51 | 0.01 | 0.01 |
| 8 | 0.18 | 0.00 | 0.01 | 0.00 | 0.00 | 0.02 | 0.05 | 0.03 | 0.68 | 0.05 |
| 9 | 0.19 | 0.00 | 0.00 | 0.00 | 0.00 | 0.01 | 0.06 | 0.01 | 0.02 | 0.70 |

Figure 4. Error matrix highlighting the classes across the rows that each actual case most often gets mistaken for. The diagonal shows the correct proportion.

Like the traditional handwriting recognition task (MNIST), we anticipate the virus-hunting community can rapidly adapt this dataset to score against a suite of algorithms, including tree-based classifiers, support vectors, and linear or logistical models. For brevity, we will include our algorithm survey in further work. As a benchmark using deep learning methods, we find an overall 80% accuracy for virus identification by families when beneware is included. We also find that once a positive malware detection occurs (by signature or heuristics), the projection of the first 1024 bytes into an image can classify with 87% accuracy the type of virus. Given the 20:1 ratio between malware and beneware, future work should add more samples than the approximately 2500 non-malicious executables included here.

## 4. DISCUSSION AND CONCLUSIONS

This short note introduces a consolidated dataset for scoring malware Portable Executable file headers when recast as an abstract image recognition problem. The image-based approach may generalize better than other heuristic methods, particularly given how virus authors change single-bytes to fool hash-based signatures but CNN detection typically proves less sensitive to small image changes. The ability to read headers only and render a readable file fingerprint as a small image suggests a potential fast detection rate seen in other vision applications (near real-time if greater than 30 frames per second by convention). Further enhancements may improve accuracy with data augmentation [28] strategies. While the reduction of file headers to images may initially seem counterintuitive, the abstract representation of text or audio into images has benefited other machine learning applications. We anticipate that further work may render simple, standalone appliances that may operate offline for successful detection, particularly for the small, fast, and accurate models like MobileNetV2. The Virus-MNIST dataset is available on Kaggle and Github [29].

## ACKNOWLEDGMENTS

The author would like to thank the PeopleTec Technical Fellows program for encouragement and project assistance.